\documentstyle[epsf,11pt]{article}
\newlength{\dinwidth}               \newlength{\dinmargin}
\begin{document}
\newcommand{\ks}       {\mbox{$ K^0_S$}}
\newcommand{\kl}       {\mbox{$ K^0_L$}}
\newcommand{\k}        {\mbox{$ K^0$}}
\newcommand{\kb}       {\mbox{$ \overline{K^0}$}}
\newcommand{\lam}      {\mbox{$\Lambda$}}
\newcommand{\lb}       {\mbox{$\overline{\Lambda}$}}
\newcommand{\GeV}      {\mbox{\rm GeV}}
\newcommand{\GeVc}     {\mbox{\rm GeV/c}}
\newcommand{\mpipi}    {\mbox{${M_{\pi\pi}}$}}
\newcommand{\mppi}     {\mbox{${M_{p\pi}}$}}
\newcommand{\mee}      {\mbox{${M_{ee}}$}}
\newcommand{\delz}     {\mbox{$\bigtriangleup Z$}}
\newcommand{\angxy}    {\mbox{$\alpha_{{\small XY}}$}}
\newcommand{\ctau}     {\mbox{$\rm c\tau$}}
\newcommand{\ra}       {\mbox{$\rightarrow $}}
\newcommand{\ptr}      {\mbox{$p_{\rm T}$}}
\newcommand{\pms}      {\mbox{$\pm$}}
\newcommand{\etamx}    {\mbox{$\eta_{\rm max}$}}
\newcommand{\pspu}    {\mbox{$P_s/P_u$}}
\newcommand{\psupudpspd}   {\mbox{$(P_{su}/P_{ud})/(P_s/P_d)$}}
\newcommand{\pqqpq}    {\mbox{$P_{qq}\P_q$}}
\vspace{1cm}
\title {
{\bf
  Neutral strange particle production in
                  deep inelastic scattering at HERA  } \\
 }
\author{\rm  ZEUS Collaboration }
\date{   }
\maketitle
\vspace{5 cm}

\begin{abstract}
This paper presents measurements of \k\ and \lam\ production in neutral
current, deep inelastic scattering of 26.7 GeV electrons and 820 GeV protons
in the kinematic range $ 10 < Q^{2} < 640 $ GeV$^2$, $0.0003 < x < 0.01$,
and $y > 0.04$. Average multiplicities for \k\ and \lam\ production are
determined for transverse momenta \ \ptr\ $> 0.5 $ GeV and pseudorapidities
$\left| \eta \right| < 1.3$.  The multiplicities favour a stronger strange
to light quark suppression in the fragmentation chain than found in $e^+
e^-$ experiments.  The production properties of \k's in events with and
without a large rapidity gap with respect to the proton direction are
compared. The ratio of neutral \k's to charged particles per event in the
measured kinematic range is, within the present statistics, the same in both
samples.
\end{abstract}

\vspace{-16cm}
\begin{flushleft}
\tt DESY 95-084 \\
April 1995 \\
\end{flushleft}

\setcounter{page}{0}
\thispagestyle{empty}

\newpage

\def\3{\ss}
\parindent 0cm
\footnotesize
\renewcommand{\thepage}{\Roman{page}}
\begin{center}
\begin{large}
The ZEUS Collaboration
\end{large}
\end{center}
M.~Derrick, D.~Krakauer, S.~Magill, D.~Mikunas, B.~Musgrave,
J.~Repond, R.~Stanek, R.L.~Talaga, H.~Zhang \\
{\it Argonne National Laboratory, Argonne, IL, USA}~$^{p}$\\[6pt]
R.~Ayad$^1$, G.~Bari, M.~Basile,
L.~Bellagamba, D.~Boscherini, A.~Bruni, G.~Bruni, P.~Bruni, G.~Cara
Romeo, G.~Castellini$^{2}$, M.~Chiarini,
L.~Cifarelli$^{3}$, F.~Cindolo, A.~Contin, M.~Corradi,
I.~Gialas$^{4}$,
P.~Giusti, G.~Iacobucci, G.~Laurenti, G.~Levi, A.~Margotti,
T.~Massam, R.~Nania, C.~Nemoz, \\
F.~Palmonari, A.~Polini, G.~Sartorelli, R.~Timellini, Y.~Zamora
Garcia$^{1}$,
A.~Zichichi \\
{\it University and INFN Bologna, Bologna, Italy}~$^{f}$
% ...
\\[6pt]
\clearpage
\hspace*{1mm}
$^{ 1}$ supported by Worldlab, Lausanne, Switzerland \\
\hspace*{1mm}
$^{ 2}$ also at IROE Florence, Italy  \\
\hspace*{1mm}
$^{ 3}$ now at Univ. of Salerno and INFN Napoli, Italy  \\
\hspace*{1mm}
$^{ 4}$ supported by EU HCM contract ERB-CHRX-CT93-0376 \\
\hspace*{1mm}
% ...
$^{18}$ now with OPAL Collaboration, Faculty of Physics at Univ. of
        Freiburg \\
% ...

\begin{tabular}{lp{15cm}}
$^{a}$ &supported by the Natural Sciences and Engineering Research
         Council of Canada (NSERC) \\
$^{b}$ &supported by the FCAR of Qu\'ebec, Canada\\
$^{c}$ &supported by the German Federal Ministry for Research and
         Technology (BMFT)\\
$^{d}$ &supported by the MINERVA Gesellschaft f\"ur Forschung GmbH,
         and by the Israel Academy of Science \\
$^{e}$ &supported by the German Israeli Foundation, and
         by the Israel Academy of Science \\
$^{f}$ &supported by the Italian National Institute for Nuclear Physics
         (INFN) \\
$^{g}$ &supported by the Japanese Ministry of Education, Science and
         Culture (the Monbusho)
         and its grants for Scientific Research\\
$^{h}$ &supported by the Korean Ministry of Education and Korea Science
         and Engineering Foundation \\
$^{i}$ &supported by the Netherlands Foundation for Research on Matter
         (FOM)\\
$^{j}$ &supported by the Polish State Committee for Scientific Research
         (grant No. SPB/P3/202/93) and the Foundation for Polish-
         German Collaboration (proj. No. 506/92) \\
$^{k}$ &supported by the Polish State Committee for Scientific
         Research (grant No. PB 861/2/91 and No. 2 2372 9102,
         grant No. PB 2 2376 9102 and No. PB 2 0092 9101) \\
$^{l}$ &partially supported by the German Federal Ministry for
         Research and Technology (BMFT) \\
$^{m}$ &supported by the German Federal Ministry for Research and
         Technology (BMFT), the Volkswagen Foundation, and the Deutsche
         Forschungsgemeinschaft \\
$^{n}$ &supported by the Spanish Ministry of Education and Science
         through funds provided by CICYT \\
$^{o}$ &supported by the Particle Physics and Astronomy Research
        Council \\
$^{p}$ &supported by the US Department of Energy \\
$^{q}$ &supported by the US National Science Foundation
\end{tabular}

\newpage
\pagenumbering{arabic}
\setcounter{page}{1}
\normalsize

\section{\bf Introduction}
\label{s:intro}

The investigation of strange particle production in neutral current,
deep inelastic scattering (DIS) interactions could provide information
about the $s$-quarks in
the nucleon, about the boson-gluon fusion process and, above all, the parton
fragmentation process.
Strange particle production has been measured previously by
experiments where the $\gamma^* p$ centre-of-mass energy, $W$, is
at least one order of magnitude lower than at HERA
 \cite{BAKER,AMMOSOV,WA59,WA21b}.
The ratio of strange particle to light non-strange particle
production of approximately 1:5 is ascribed to a reduced
probability of strange quark creation in
the parton fragmentation chain. In simulation programs
based on the Lund scheme it is parametrised by
the strange quark suppression factor $P_s/P_u$.
Here $P_s$ and $P_u$ are the probabilities for creating
$s-$ or $u,d-$quarks from the vacuum during the fragmentation process.
A detailed review of our knowledge on heavy quark suppression is
given in \cite{WRO}.
In hadron-hadron collisions an increasing $P_s/P_u$ is found with
increasing centre-of-mass energy. Also indications of a dependence of
the strangeness suppression factor on the region of phase space under
investigation are reported.
The values found vary between about 0.15 and 0.55 with a mean value
close to 0.3 (see for example \cite{E665,PREVIOUS}).
The parameters for the hadronisation process in the present day electron-proton
Monte Carlo event generators are obtained from fits to $e^+e^-$ data
and are assumed to be the same in DIS experiments due to jet universality.

The longitudinal phase space of the $\gamma^* p$ interactions
at HERA can be divided into three
regions where different processes are expected to dominate. These processes
also appear in $e^+e^-$ or hadron-hadron scattering:
a) the fragmentation region of the struck quark, which resembles that
of one of the pair-produced quarks in $e^+e^-$
annihilation
experiments; b) the fragmentation of the proton remnant, which resembles
the fragmentation  in hadron colliders; and c) the hadronic centre-of-mass
central rapidity  region, where the colour flow between the struck quark
and the proton remnant evolves.
The latter region exists in both $e^+e^-$ and hadron collider
experiments.
The acceptance of our central tracking detector allows us to study
\k\ production in the fragmentation region of the struck quark
and the central rapidity region.
The part of the event which is well inside our detector
acceptance is dominated by particles originating from the central
rapidity region.

In about 10\% of the DIS events no
proton remnant is detected in the ZEUS detector,
resulting in a large rapidity gap (LRG)
between the acceptance limit in the proton direction and the first
visible particle in the detector
\cite{JETLRG,H1LRG}. The properties of these events are consistent
with the assumption that the exchanged photon is scattered off a colourless
object emitted by the proton. This object is
generically called a pomeron.
There exist indications that the pomeron has a partonic substructure
\cite{UA8,JETLRG} but the nature of its constituents is still under
investigation. A natural assumption is that they are quarks
and gluons or a combination of both.
It is expected that the strange quark content of the
pomeron could affect the strange particle multiplicity in the final state
of these events.

The investigation of strange particle production allows us to connect
results from
$e^+e^-$ experiments and from hadron collider experiments.
This paper is a first step of such a program. We compare the \k\ and
\lam\ multiplicities\footnote{Throughout this paper, a reference
         to a particle includes a reference to its antiparticle.}
and their momentum and angular distributions in the new kinematic
region of HERA with extrapolations from Monte Carlo models
based on the results of lower
energy experiments.
The $Q^2$ evolution of the \k\ multiplicity is studied.
The production of \k 's in events with a large rapidity gap is compared
to that of events without a large rapidity gap.

All studies are performed in the HERA laboratory frame and are restricted to a
kinematic range where the tracking  acceptance is high and well understood.

\section{\bf Experimental setup}
\subsection*{\bf HERA machine conditions}

The data were collected at the electron-proton collider HERA
using the ZEUS detector during the 1993 running period. HERA collided
26.7~GeV electrons with 820~GeV protons.
84 bunches were filled for each beam and in addition 10 electron
and 6 proton bunches were left unpaired for background studies.
The typical electron and proton currents were 10~mA leading to a
typical instantaneous luminosity of $6 \cdot 10^{29}~{\rm cm}^{-2}
{\rm s}^{-1}$.
An integrated luminosity of 0.55~pb$^{-1}$ of data was collected in 1993.

\subsection*{\bf The ZEUS detector}
ZEUS is a multipurpose, magnetic detector which
has been described elsewhere \cite{ZEUS1}. Here we give
a brief description concentrating on those parts of the detector
relevant for the present analysis.

Charged particles are tracked by the inner tracking detectors which
operate in a magnetic field of 1.43 T provided by a thin superconducting
solenoid surrounding the tracking detectors.
Immediately after the beampipe there is a cylindrical drift
chamber, the vertex detector (VXD),
which consists of 120 radial cells, each with 12 sense wires \cite{VXD}.
The achieved resolution is 50~$\mu$m in the central region of a cell
and 150~$\mu$m near the edges. Surrounding the VXD is the central
tracking detector (CTD) which consists of 72 cylindrical drift chamber
layers, organised into 9 ``superlayers'' \cite{CTD}. Each superlayer
consists either of wires parallel (axial) to the beam
axis or of wires inclined at a small angle to give a stereo view.
With the present understanding
of the chamber, a spatial resolution of 260~$\mu$m has been achieved.
The hit efficiency of the chamber is greater than 95\%.

In events with charged particle tracks, using the combined data from
both chambers, reconstructed primary vertex position resolutions of
0.6~cm in the $Z$ direction and 0.1~cm in the $XY$ plane are
measured\footnote{The ZEUS coordinate system is defined as right-handed
with the $Z$ axis pointing in the proton beam direction and the $X$ axis
horizontal pointing towards the centre of HERA. The polar angle $\theta$
is defined with respect to the $Z$-direction.}.
The resolution in transverse momentum for full length tracks is
$\sigma(p_{\rm T}) / p_{\rm T} =
\sqrt{ (0.005~ p_{\rm T})^2 + (0.016)^2} $ ($p_{\rm T}$ in GeV).

The solenoid is surrounded by a high resolution uranium-scintillator
calorimeter divided into three parts, forward (FCAL), barrel (BCAL)
and rear (RCAL). Holes of 20 $\times$ 20~cm$^2$ in the centre of FCAL and
RCAL are required to accommodate the HERA beam pipe. Each of the
calorimeter parts is subdivided into towers which in turn are segmented
longitudinally into electromagnetic (EMC) and hadronic (HAC) sections.
A section of a tower is called a cell and is read out by
two photomultiplier tubes. A detailed description of the calorimeter
is given in \cite{CALRes}.

For measuring the luminosity via the Bethe-Heitler process $ e p \rightarrow
e' p' \gamma$, as well as for tagging very small
$Q^2$ processes, two lead-scintillator calorimeters are used
\cite{LUMIhard}. Bremsstrahlung photons emerging from the
electron-proton interaction region at angles
$\theta'_{\gamma}\le$ 0.5~mrad with respect to the electron beam
axis hit the photon calorimeter at 107~m from the interaction point
(IP). Electrons
emitted from the IP at scattering angles less than or equal to 6~mrad
and with energies  between
20\% and 90\% of the incident electron energy are deflected
by beam magnets and hit the electron calorimeter placed 35~m from
the IP.

\section{\bf HERA kinematics}

The kinematics of deep inelastic scattering processes at HERA,
$e^- p \rightarrow e^- h$,
where $h$\ is the hadronic final state, can be described by the Lorentz
invariant variables $Q^2$, $x$ and $y$.
Here $-Q^2$ is the square of the four-momentum
transfer between the incoming electron and the scattered electron;
$x$, in the na\"{\i}ve quark-parton model,
is the fractional momentum of the struck quark in the proton, and $y$ is the
relative energy transfer of the electron to the hadronic system.
The variables  are related
by $ Q^2 = s x y $, where $s$\ is the squared invariant
mass of the $e p$ \ system.
$Q^2$, $x$ and $y$ can be calculated from the kinematic variables of the
scattered electron, from the hadronic final state variables, or from a
combination of both.
The optimal reconstruction method depends on the event kinematics and
the detector resolution.

In this paper we use the double angle method \cite{Q2DA}
to calculate the $Q^2$ and $x$ variables:
\begin{eqnarray*}
Q^2_{DA} & = &4 E_e^2 \cdot \ { \sin \gamma_h \ (1+ \cos \theta_e) \over
 \sin \gamma_h + \sin \theta_e - \sin (\gamma_h + \theta_e)}, \\
x_{DA} & = &{E_e \over E_p} \cdot
 { \sin \gamma_h + \sin \theta_e + \sin(\gamma_h + \theta_e) \over
 \sin \gamma_h + \sin \theta_e - \sin (\gamma_h + \theta_e)}.
\end{eqnarray*}
Here $E_e$ and $E_p$ are the initial electron and proton energies;
$\theta_e$ is the electron scattering angle with respect to the
incident proton direction and
$\gamma_h$ is the polar angle of a massless object balancing
the momentum vector of the scattered electron to satisfy four-momentum
conservation.  In the na\"{\i}ve quark-parton model $\gamma_h$ is the
scattering angle of the struck quark.
It is determined from the hadronic energy flow in the calorimeter:
$$\cos \gamma_h =
{(\sum p_X)_h^2 + (\sum p_Y)_h^2 - (\sum E-p_Z)_h^2
\over
(\sum p_X)_h^2 + (\sum p_Y)_h^2 + (\sum E-p_Z)_h^2}.$$
Here the sums run over all calorimeter cells which are not
assigned to the scattered electron and ($p_X,p_Y,p_Z$) is the momentum vector
assigned to each cell of energy $E$. The cell angles are calculated from the
geometric centres of the cells and the vertex position of the event.

Using the hadronic energy flow of the final state,
$y$\ can be calculated
according to the Jacquet-Blondel method \cite{JacBlon}:
$$y_{JB} = {1 \over 2 E_e } \ \sum_h \ (E-p_{Z})_h. $$

For background rejection we also
calculate $y$ using the electron information:
$$ y_e  =  1 -  {E^{\prime}_e \over 2 E_e} \ (1-\cos \theta_e),$$
where $E'_e$ is the energy of the scattered electron.
The square of the centre-of-mass energy of the
virtual photon-proton system, $\gamma^* p$, is calculated using:
$$ W^2_{DA} = m_p^2 + Q^2_{DA} ({1 \over x_{DA}} - 1), $$
where $m_p$ is the proton mass. We use the described  methods for
calculating the kinematic variables and do not mention them explicitly
below except when necessary.

\section{\bf Event selection}
\subsection{\bf Trigger conditions}
The trigger is organised in three levels \cite{ZEUS1}.
For DIS events, the first level trigger (FLT) requires
at least one of three conditions of energy
sums in the EMC calorimeter: the BCAL EMC energy exceeds 3.4 GeV;
the RCAL EMC energy (excluding the innermost towers surrounding the beam pipe)
exceeds 2.0 GeV; or the RCAL EMC energy (including those towers) exceeds
3.75 GeV.

The second level trigger (SLT) rejects proton beam-gas events by
using the  times measured in the calorimeter cells.
The DIS trigger rate
of the SLT is about one-tenth of the FLT DIS trigger rate.
The loss of DIS events at the SLT is negligible.

The third level trigger (TLT) has the full event information available and
uses physics-based filters. It applies tighter timing cuts to suppress
beam-gas background further and also rejects beam halo muons and cosmic muons.
The TLT selects DIS event candidates by calculating:
$$\delta~ =~ \sum_i E_i\cdot (1-\cos\theta_i)~~~>
         ~~~ 20~{\rm GeV} ~~ - ~~2~E_{\gamma},$$
where $E_i$ and $\theta_i$ are the energy and the polar angle of
the energy deposits in the central calorimeter.
$E_{\gamma}$ is the
energy measured in the photon calorimeter of the luminosity monitor.
The summation runs over all energy deposits in the calorimeter cells.
For fully contained DIS events $\delta \approx 2 E_e = 53.4$ GeV.
Photoproduction events have low values
of $\delta$ compared to DIS events because the scattered electron
escapes in the beam pipe hole of the rear calorimeter.
For events with $Q^2$ \  less than $\sim$~4~GeV$^2$
the calorimeter cannot detect the scattered electron.

For events with the scattered electron detected in the calorimeter,
the trigger acceptance was essentially independent of the DIS hadronic
final state. It was greater than 97\% for $Q^2>10$~GeV$^2$
and independent of $Q^2$.
A total of $7 \cdot 10^6$ \ events passed the TLT and
was written to tape during the 1993 running period.

\subsection{Offline event selection}

The offline selection of DIS events is similar to
that described in our earlier publication \cite{F2}.
The characteristic signature of a DIS event is the scattered electron
detected in the uranium scintillator calorimeter. The pattern of energy
deposition in the calorimeter
cells is used to identify an electron candidate.
We use the following criteria to select a sample of DIS
events:
\begin{itemize}
\item a scattered electron candidate has to be found with
      $E^{\prime}_e > 5 $ GeV
      and an impact point at the RCAL surface outside a square
      of 32 x 32 cm$^2$ centred on the beam line.
      This requirement ensures that the electromagnetic shower is
      fully contained
      within the calorimeter and its impact point can be reconstructed
      with sufficient accuracy;
\item $ y_{e} < 0.95 $  \ \ to reduce  photoproduction background;
\item  35~GeV $ < \delta < 60 $ GeV  \ \ to remove
        photoproduction events and to
        suppress events with hard initial state radiation;
\item --50~cm $< Z < 40$~cm, \ \ where $Z$ is the position of
       the  event vertex reconstructed from the CTD.
       This requirement rejects beam-gas and cosmic ray events.
\end{itemize}

  From Monte Carlo studies we find an average electron finding efficiency
  of 95\% in the kinematic range considered, being above 98\% in most of
  the kinematic range and dropping below 70\% for high $y$ events. The
  purity is better than 96\% for electron energies above 10~GeV and drops
  to about 60\% at high $y$.
  A total of 91000 events survive these criteria.

The particle multiplicity  and the kinematics of particle production
depend on $Q^2$
and $x$. We have restricted our analysis to a kinematic
range in $Q^2,x$ and $y$ in which migration effects are small
\cite{F2} and have little influence on the momentum and angular distributions
of the \k 's and \lam 's. We chose the following range:
\begin{itemize}
\item $ 10~{\rm GeV}^2 < Q^2 < 640 $ GeV$^2$;
\item $ 0.0003 < x < 0.01$;
\item $ y > 0.04.$
\end{itemize}

The $Q^2$ and $x$ variables are calculated according
to the double angle method and $y$ with the Jacquet-Blondel method.
After applying these criteria to the previously selected sample we are left
with 27500 events.

\section{Monte Carlo simulation}

Monte Carlo event simulation is used to determine the acceptance and
resolution of the ZEUS detector. The simulation is based on the GEANT 3.13
\cite{GEANT} program and incorporates the knowledge of the detector
and the trigger.

\subsection*{Simulation of normal DIS events}

Neutral current DIS events with $Q^2>4$ GeV$^2$ were generated using the
HERACLES 4.4 program \cite{SPIES} which incorporates first order
electroweak corrections. The Monte Carlo program LEPTO 6.1 \cite{INGEL},
interfaced to HERACLES via the program DJANGO 6.0 \cite{DJANGO}, was used to
simulate QCD cascades and fragmentation. The parton cascade was modelled
in different ways:

\begin{itemize}
\item
  the colour-dipole model including the boson-gluon fusion
  process (CDM) as implemented in the ARIADNE 4.03 \cite{ARI} program
  was used. In this model coherence effects are implicitly included in
  the formalism of the parton cascade;
\item
  matrix element calculations plus the parton shower option
  (MEPS) as implemented in
  LEPTO were used, where coherence effects in the final state cascade are
  included by angular ordering of successive parton emissions.
\end{itemize}
These models use the Lund string fragmentation \cite{LUND} for the
hadronisation phase as implemented in JETSET 7.3 \cite{JET}.

For the CDM event sample the
MRSD$^{\prime}_{-}$ parton density parametrisation for the
proton was used \cite{MRSD}. The GRV \cite{GRV} parametrisation
was used for the MEPS data set. These parametrisations describe reasonably
the HERA measurements of the proton structure function
F$_2$ \cite{Z93F2,H93F2}.

The simulations predict that about 10\% of the \ks 's are produced in
charm events and about 5\% originate from sea quarks in the
proton.
The remaining $\sim 85$\% of the \ks 's are created in the fragmentation
chain depending on the actual value of the strange-quark suppression
factor $P_s/P_u$.
The parameters of the Monte Carlo models are set to their default values
($P_s/P_u = 0.3$).
We have also generated events with
$P_s/P_u = 0.2$ as suggested in \cite{E665} and we have compared the
predictions of the simulations with the measured rates.
Since the MEPS model and the CDM model behave similarly when reducing
the $P_s/P_u$ parameter, we only show the predictions with $P_s/P_u = 0.2$
for the CDM model.

\subsection*{Simulation of large rapidity gap DIS events}

Our previous study \cite{JETLRG} shows that diffractive models,
specifically POMPYT \cite{POMPYT} and a model by  Nikolaev and
Zakharov \cite{NZ} as implemented in our Monte Carlo
program NZ \cite{ADA},
give adequate descriptions of the properties of the
LRG events. We have used POMPYT and NZ event samples for our study
of \k\ multiplicities in events with a large rapidity gap.
The POMPYT Monte Carlo program uses an implementation of the
Ingelman and Schlein model \cite{IS}, describing high energy diffractive
processes.
In this model the virtual photon interacts with
the constituents of the pomeron emitted by the proton.
Factorisation is assumed in the sense that the pomeron emission
and the pomeron structure are independent.
The current version of POMPYT contains no strange quark constituents
for the pomeron.
The NZ Monte Carlo model on the other hand is non-factorisable.
Here the virtual photon fluctuates into a $q\overline {q}$
or a $q\overline {q} g$ state and interacts with a colourless two-gluon
system emitted by the proton.
The $q \overline q g$ states were fragmented as if they were $q \overline q$
states and the flavours are generated in 90\% of the cases as ($u, d$) and
in 10\% as $s$.

\section{\bf Selection of \ks\ and \lam\ candidates}

\ks\ particles are identified in the decay channel
$\ks \rightarrow \pi^+ \pi^- $ and \lam\ particles are detected in
the channel
\lam\ $\rightarrow p \pi^-$.
Due to their lifetime of ${\cal O}$(10$^{-10}$s) and their typical
momenta of about 1~GeV they have an average decay
length of a few centimetres, which results in secondary vertices well
separated from the primary event vertex.

Tracks are reconstructed using the CTD and the VXD. The track finding
algorithm starts with hits in the outermost axial superlayers of the
CTD. As the trajectory is followed inwards to the beam axis, more hits
on the axial wires and from the VXD are assigned to the track.
The resulting circle in the
transverse plane is used for the pattern recognition in the stereo
superlayers.  The momentum is determined in a 5-parameter helix
fit. Multiple Coulomb scattering in the beam pipe and in the outer walls
of the VXD is taken into account in the evaluation of the covariance
matrix.

The primary event vertex is determined from a $\chi^2$ fit performed with
the tracks using the perigee parametrisation \cite{SIJIN}
and assuming that the tracks come from a common point in space.
A track is considered not to be associated with
the primary vertex if the $\chi^2$ for the primary vertex fit increases
significantly when the track is included in the fit.

The systematic effects in the CTD are most serious for
low $p_{\rm T}$ tracks and for tracks which
traverse the inhomogeneous part of the magnetic field at the
ends of the CTD. The reconstructed tracks used in this analysis were
required to have a transverse momentum \ptr\ $>$ 0.2 GeV and a polar angle
between $25^\circ < \theta < 155^\circ$.
In terms of pseudorapidity, $\eta = -\log (\tan (\theta / 2))$, this
corresponds to $\left| \eta \right| < 1.5 $.
This is the region where the CTD response and systematics are well understood.

\subsection{ \bf \ks\ Identification}

To search for \ks, we examine pairs of oppositely charged tracks to find
a secondary vertex.
We refer to these tracks as daughter tracks. At least one of
the daughter tracks is not allowed to be associated with the primary vertex
and track pairs which do not intersect when projected into
the transverse plane are rejected.

For each remaining track pair, we
obtain the momentum of the \ks\ candidate by calculating the momenta of the
individual tracks at their intersection point and adding them.
\ks\ candidates with transverse momenta below  0.5~GeV or above 4~GeV or
with directions of flight too near to the beam pipe,
$\left| \eta \right| > 1.3$, are removed.

\begin{figure}[tp]
\includegraphics{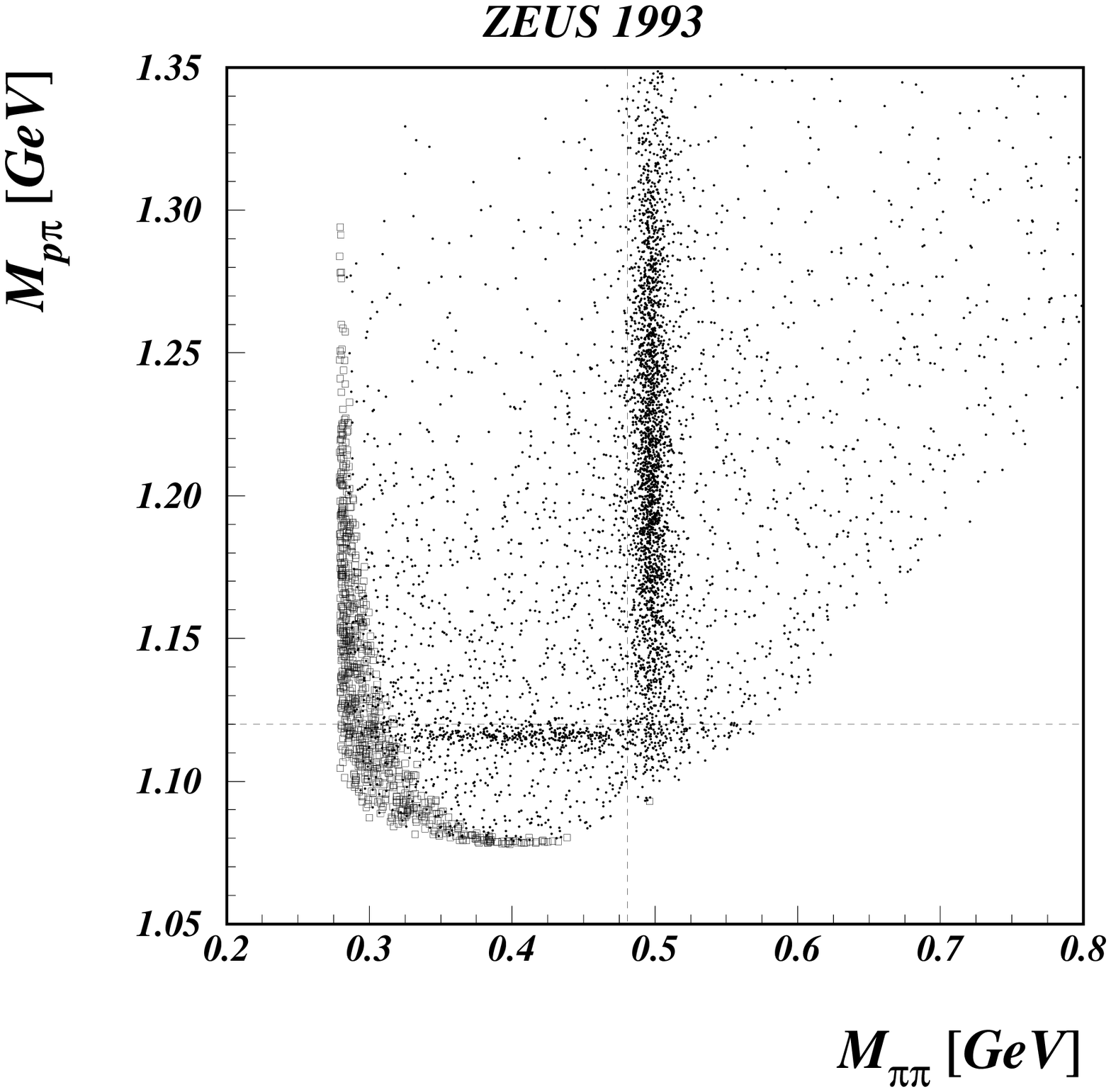}
\unitlength1cm
\begin{picture}(20,20)
\thicklines
\end{picture}
\caption{  Invariant masses for two particle combinations
assuming a $p \pi$ versus a $\pi \pi$ mass
hypothesis.  The background to the \ks\ signal is reduced by removing the
candidates with $M_{p\pi} < 1.12$~GeV.  The background to the \lam\ signal is
reduced by removing candidates with $M_{\pi\pi} > 0.481$~GeV.
The dashed lines correspond to these mass values.
Candidates for photon conversions removed by
requiring $M_{ee} < 50$~MeV are indicated
by squares.}
\label{fig:k0_lam_scatter}
\end{figure}

The background in the mass region of the \ks\ is reduced by applying the
following criteria:
\begin{itemize}
\item
  $\cos(\alpha_{XY}) > 0.99$, \ where  $\alpha_{XY}$ is the angle in the
  transverse plane between the direction of flight of the \ks\ candidate
  and its reconstructed momentum direction;
\item
   the separation in Z between the two tracks at their
   $XY$ intersection point has to be $ \left| \Delta Z \right|
   < 2.5$~cm. The coordinates of the \ks\ decay vertex are set to
   the $XY$ coordinates of the intersection point of the track circles
   and the $Z$ coordinate is chosen to be in the centre between the
   closest approaches in $Z$ of the two track circles;
\item
   the proper lifetime of the candidates,
   $ c\tau  = (L M c) / p$,
   has to be less than 10 cm. Here $L$ \ is the decay
   length, $p\ $ is the momentum and $M$ is the invariant mass of the
   candidate;
\item
   to reduce background arising from photon conversions into $e^+e^-$ pairs,
   pairs of tracks considered as electrons must have an effective
   mass \mee$>50$ MeV (see Fig.~\ref{fig:k0_lam_scatter});
\item
   to eliminate \lam\ contamination of the \ks\ signal,
   candidates with a mass hypothesis $ \mppi < 1.12$~GeV are rejected
   (see Fig.~\ref{fig:k0_lam_scatter}).
\end{itemize}

\begin{figure}[bp]
\includegraphics{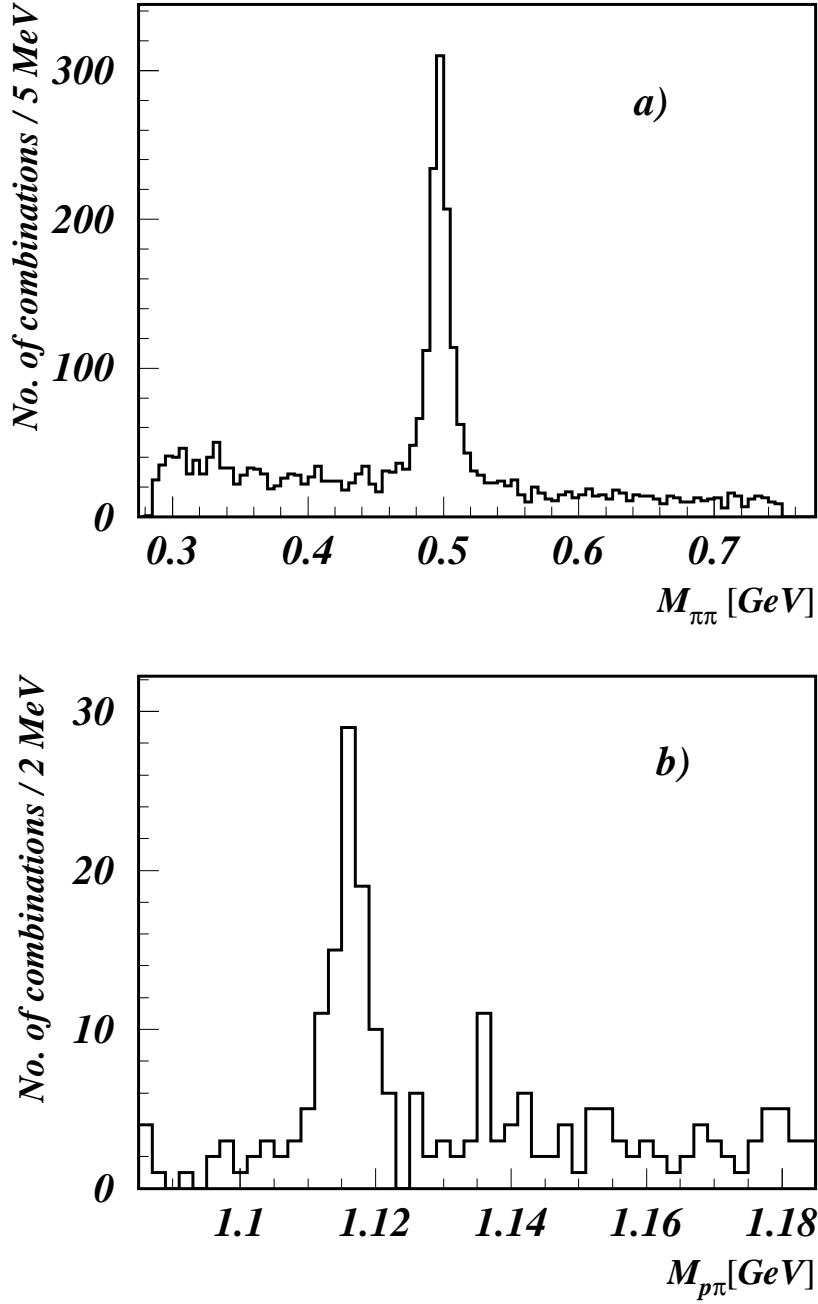}
\unitlength1cm
\begin{picture}(20,20)
\thicklines
\end{picture}
\caption{   a) Reconstructed \ks\ mass distribution and b) reconstructed
\lam\ mass distribution
in accepted events with $10 < Q^2 < 640$~GeV$^2$, $0.0003 < x < 0.01$,
and $y > 0.04$. The kinematic range of accepted \ks\
candidates is 0.5~$ < p_{\rm T} < 4$~GeV and for \lam\ candidates it is
$0.5 < p_{\rm T} < 3.5$~GeV. The $\eta$ range of accepted \ks\ and \lam\
candidates is $\left| \eta \right| < 1.3$.}
\label{fig:k0_lam_signal}
\end{figure}

Using these criteria (summarised in Tab.~\ref{tab:reconstruction_cuts})
we obtain the \ks\ signal shown in
Fig.~\ref{fig:k0_lam_signal}a.
We fit the $\pi^+ \pi^-$ mass spectrum with a Gaussian
and a linear background in the region 0.4 to 0.6 GeV.
The fitted mass is 497.4~\pms~0.3 MeV and the standard deviation
is 7.8~\pms~0.3~MeV.
The mass value and width of the signal are well reproduced by the Monte Carlo
simulations. In the signal region we find a total of
971 \ks\ mesons on top of a background of about 150 $\pi\pi$-combinations.
The \ks\ signal region extends from 474 to 521 MeV.
The average lifetime of the \ks\ mesons was determined by fitting
the exponential form $\exp(-c\tau / c\tau_{\ks})$ to the acceptance corrected
$c\tau$ lifetime distribution. Here the $c\tau$ upper limit was relaxed to
20~cm and all other selection criteria were set to their default value.
The result $c\tau_{\ks } = 2.66 \pm 0.11 \pm 0.06$~cm
is consistent with the world average of 2.676 $\pm$ 0.006 cm given in
\cite{PDG}.
The systematic uncertainty includes the variation of the number of bins used in
the fit and tightening or loosening the selection criteria.

\subsection{ \bf \lam\ Identification}

The \lam\ identification closely resembles the \ks\ identification.
The daughter track with the higher momentum is considered to be the proton.
No daughter track is allowed to be associated with the primary vertex.
The \ $c\tau\ $ upper limit is increased to 40~cm in order to account
for the longer lifetime of the \lam.
Requiring $M_{\pi\pi} < 0.481 $~GeV  removes
the background from \ks\ mesons (see Fig.~\ref{fig:k0_lam_scatter}).
Since there is no clear \lam\ signal seen for candidates with \ptr\
above 3.5~GeV, this value is chosen as the upper limit of the investigated
momentum range.

Figure~\ref{fig:k0_lam_signal}b shows the \lam\ signal obtained.
We fit the $p \pi$~mass spectrum from 1085 to 1185 MeV.
The fit yields a mass of 1116.2~\pms~0.4~MeV with a standard deviation
of 3.0~\pms~0.5~MeV.
The Monte Carlo simulation reproduces well the \lam\ mass position and width.
Within the signal region we find 80 \lam\ baryons and 18 background
combinations. The signal region runs from 1107 to 1125~MeV.
Of the 80 \lam\ baryons, (60 $\pm$ 5)\% are \lb\  and the remaining
(40 $\pm$ 5)\% are \lam.
The determination of the average lifetime of the \lam\ from
the lifetime distribution gives $c\tau_{\lam } = 7.3 \pm 2.2 \pm 0.5$~cm,
consistent with the value of 7.89~cm given in \cite{PDG}.

\begin{table}[htb]
\begin{center}
\begin{tabular}{|l|c|c|}
\hline
Selection parameters for candidates & \ks           & \lam       \\
\hline
$\cos(\angxy)$                      & $>$~0.99      & $>$~0.99   \\
$\left| \delz \right|$  [cm]        & $<$~2.5       & $<$~2.5    \\
\ctau\ [cm]                         & $<$~10        & $<$~40     \\
\mppi\ [GeV]                        & $>$~1.12      &  -         \\
\mpipi\ [GeV]                       &  -            & $<$~0.481  \\
\mee\ [GeV]                         & $>$~0.05      & $>$~0.05   \\
\ptr$_{\ daughter-tracks}$ [GeV]    & $>$~0.2       & $>$~0.2    \\
$\theta_{daughter-tracks}$ [$^o$]   & [25,~155]     & [25,~155]   \\
No. of tracks from primary vertex   & $\le$~1       & 0          \\
\hline
$\eta$  range                       & [--1.3,~1.3]  & [--1.3,~1.3]\\
\ptr\  [GeV] range                  & [0.5,~4.0]    & [0.5,~3.5]  \\
\hline
\end{tabular}
\caption{
Selection criteria for \ks\ and \lam\ identification.}
\label{tab:reconstruction_cuts}
\end{center}
\end{table}

\section{\bf Data correction}

   This analysis uses two types of selection criteria.  The first
   kind is event based and selects a reasonably pure sample of DIS
   events with minimal contamination from background (photoproduction,
   beam-gas, cosmic-ray events).
   The second kind of selection criteria
   is particle based and selects  a sample of \ks\ and
   \lam\ particles from the event sample defined above.

   We find a 90\% event selection efficiency, where we define
   the efficiency as the ratio of the number of Monte Carlo events passing
   all the event selection criteria (including those that restrict the
   kinematic range in $Q^2,x$  and $y$) to the total number of generated
   events in the restricted kinematic region.

   We have restricted the \k\ and \lam\ kinematic ranges
   to regions where our systematic uncertainties are small:
   their pseudorapidity is limited to $-1.3 < \eta < 1.3$ and
   their transverse momentum is restricted to a
   \ptr\ between 0.5~GeV and 4.0~GeV (3.5~GeV) for \k 's (\lam 's).
 We do not extrapolate our results to the full \ptr\ and $\eta$ range
 in order not to be dominated by model predictions. The models are
 known to have
 uncertainties especially in the low \ptr\ region and are not yet compared
 to particle properties in the proton fragmentation region of HERA  events.

   The \ks\ and \lam \ reconstruction efficiencies  were
   determined as a function of \ptr\ and $\eta$.
   For each particle type, the efficiency in a given (\ptr, $\eta$) bin
   was defined as the ratio of the number of reconstructed particles
   in the bin to the number of generated particles in the bin.
   The $\eta$ and \ptr\ resolutions are less than 5\% of the bin width
   chosen for the plots and show no systematic shifts.
   The DIS Monte Carlo events that passed all the selection criteria were
   used for these calculations.
   The \ks\ reconstruction efficiency in the kinematic region
   considered varies between  20\% for low \ptr\ and 55\% for \ptr\ above
   1.5 GeV. The efficiency varies in $ \eta $  from
   30\% around $\eta = \pm 1.3$ to 40\% for \ks 's moving transversely
   to the beam direction ($\eta = 0$).
   The \lam\ reconstruction efficiency varies between 5\% for low
   transverse momentum and approaches 20\%  for high \ptr.
   The efficiency varies in $ \eta $ between 10\% and 15\%.
   The largest loss of true \ks 's and \lam 's results from the
   collinearity requirement ($\alpha_{XY}$) and the requirement that
   daughter tracks are unassociated with the event vertex.
   Each requirement rejects about 25\% of the candidates if no other
   selection criterion is applied.

   The \k (\lam) measurements are corrected for the above efficiencies
   as well as for the branching ratios \k\ to \ks\ and
   $\ks \rightarrow \pi^+ \pi^- \quad (\lam \rightarrow p \pi )$
   \cite{PDG}.
   No corrections were made to the measurements for migrations
   and initial state radiation effects since the
   changes predicted from Monte Carlo studies are small.
   Instead we include these
   effects in our systematic error analysis (see section 9).

   The analysis procedure was checked using the reconstructed CDM (MEPS)
   Monte Carlo events as if they were data events and correcting them with
   the efficiencies obtained with the MEPS (CDM) samples.
   The corrected Monte Carlo distributions agreed at the 5\% level with the
   generated distributions.

   For the comparison of \k\ production in events with and
   without a large rapidity gap, the two-dimensional (\ptr, $\eta$)
   efficiencies were determined from the
   standard DIS Monte Carlo events satisfying the additional requirement
   $W>140$~GeV
   (see section 8.2 for details). This corresponds to a
   restriction to $y > 0.22$.
   Both non-LRG (NRG) and LRG data samples were corrected with the
   same efficiencies.
   It has been checked that the corrected and generated \ks\ distributions
   of the LRG Monte Carlo events agree well when using the efficiencies
   of those Monte Carlo sets.

  The ratio of \k\  to charged particle multiplicity,
  N(\k )/N(tracks), is investigated below.
  The charged particle multiplicity, N(tracks), is determined
  for charged particles originating at the primary vertex
  and produced in the restricted kinematic range
  $\left| \eta \right| <1.3$ and $\ptr > 0.2$~GeV.
  The number of reconstructed tracks
  is corrected for tracking inefficiencies, wrong assignments by
  the vertex finding routine to the decay products of long lived
  particles and pair conversions by using standard Monte Carlo techniques.
  The Monte Carlo corrections for particle based selection criteria
  were below 10\%.

\section{\bf Results}
\subsection{\k\  and \lam \ multiplicity distributions}
\begin{figure}[tp]
\epsfxsize=16cm
\epsfysize=20cm
\epsfbox{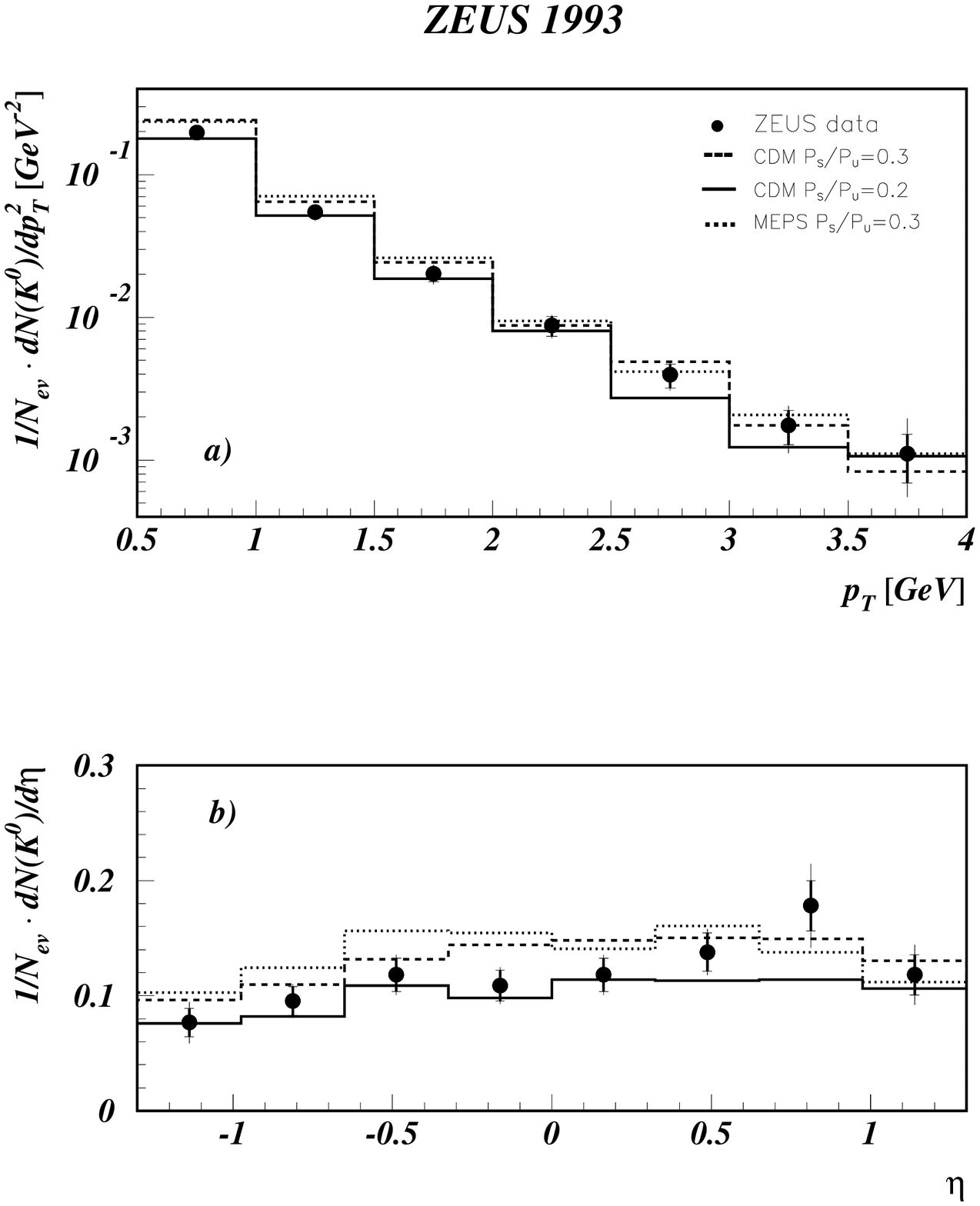}
\caption{  a) Differential multiplicity
of \k\ versus transverse momentum of the kaons in the restricted
kinematic $\eta $ \ and \ptr\ range;
b) same for the pseudorapidity of the kaons.
The inner error bars show the statistical error and the outer ones
correspond to the statistical and systematic errors added in quadrature.
The predictions of the three models discussed in the text are shown.}
\label{fig:k0_rate_pt_eta_lab}
\end{figure}

Figure~\ref{fig:k0_rate_pt_eta_lab} shows the
differential \k\ multiplicity as a function of \ptr\ and $\eta$.
The inner error bars are statistical errors and the outer ones
statistical and systematic errors added in quadrature.
The distributions are normalised by the number of events $N_{ev}$.
The predictions of the CDM and the MEPS models are overlaid.
The two curves for the CDM sample
are generated with different strange-quark suppression
factors $P_s/P_u$.
The predicted multiplicity for the default strange-quark suppression
factor of 0.3 is higher than measured.
Using the smaller suppression factor of 0.2 reduces the predicted
multiplicity to a value closer to that observed in the data. Both
parameters give a reasonable description of the measured shapes.

For events with $10<Q^2<640$~GeV$^2$, $0.0003<x<0.01$ and $y>0.04$,
the number of neutral kaons per event with $0.5<\ptr<4.0$~GeV and
$| \eta | < 1.3$ is 0.289~\pms~0.015~\pms~0.014.
The first error is statistical, the second error is systematic.
A function of the form ${C_1} / {p_{\rm T}} \cdot \exp{(C_2 p_{\rm T})}$
fits well the measured $1/N_{ev} \cdot dN(K^0)/dp^2_T$ distribution
as a function of \ptr\
over the \ptr\ range shown in
Fig.~\ref{fig:k0_rate_pt_eta_lab}.
$C_1$ and $C_2$ are constants.
The slope, $C_2$, of the \ptr\ distribution for the \k 's
is $-$1.31~\pms~0.09~\pms~0.06 GeV$^{-1}$. These values, together with the
predictions from Monte Carlo models, are listed in
Tab.~\ref{tab:kaon_results}. According to Monte Carlo studies,
the fraction of $K^0$'s produced
in the restricted \ptr \ and $\eta$ range is 23\% of the
total number of $K^0$'s produced in the final state.

\begin{table}[htbp]
\begin{center}
\begin{tabular}{|l|l|l|}
\hline
        &N(\k ) / event    & \ptr\ slope [GeV$^{-1}$] \\
\hline
  Data &0.289 \pms~0.015 \pms~0.014&--1.31 \pms~0.09 \pms~0.06 \\
CDM &   &    \\
\ \ \ with $P_s/P_u = 0.3$ & 0.342 \pms~0.005 & --1.40 \pms~0.05 \\
\ \ \ with $P_s/P_u = 0.2$ & 0.264 \pms~0.003 & --1.37 \pms~0.04 \\
MEPS &   &   \\
\ \ \ with $P_s/P_u = 0.3$ & 0.348 \pms~0.006 & --1.36 \pms~0.05 \\
\hline
\end{tabular}
\caption{
Results of the \k\ measurement for events with $10<Q^2<640$~GeV$^2$,
$0.0003<x<0.01$, $y>0.04$ and for a \k\ with $0.5 < \ptr < 4.0$~GeV and
 $| \eta | <1.3$.
The two CDM samples have been generated with a different strange-quark
suppression factor  $P_s/P_u$.
\label{tab:kaon_results}
}
\end{center}
\end{table}

\begin{figure}[htbp]
\epsfxsize=16cm
\epsfysize=20cm
\epsfbox{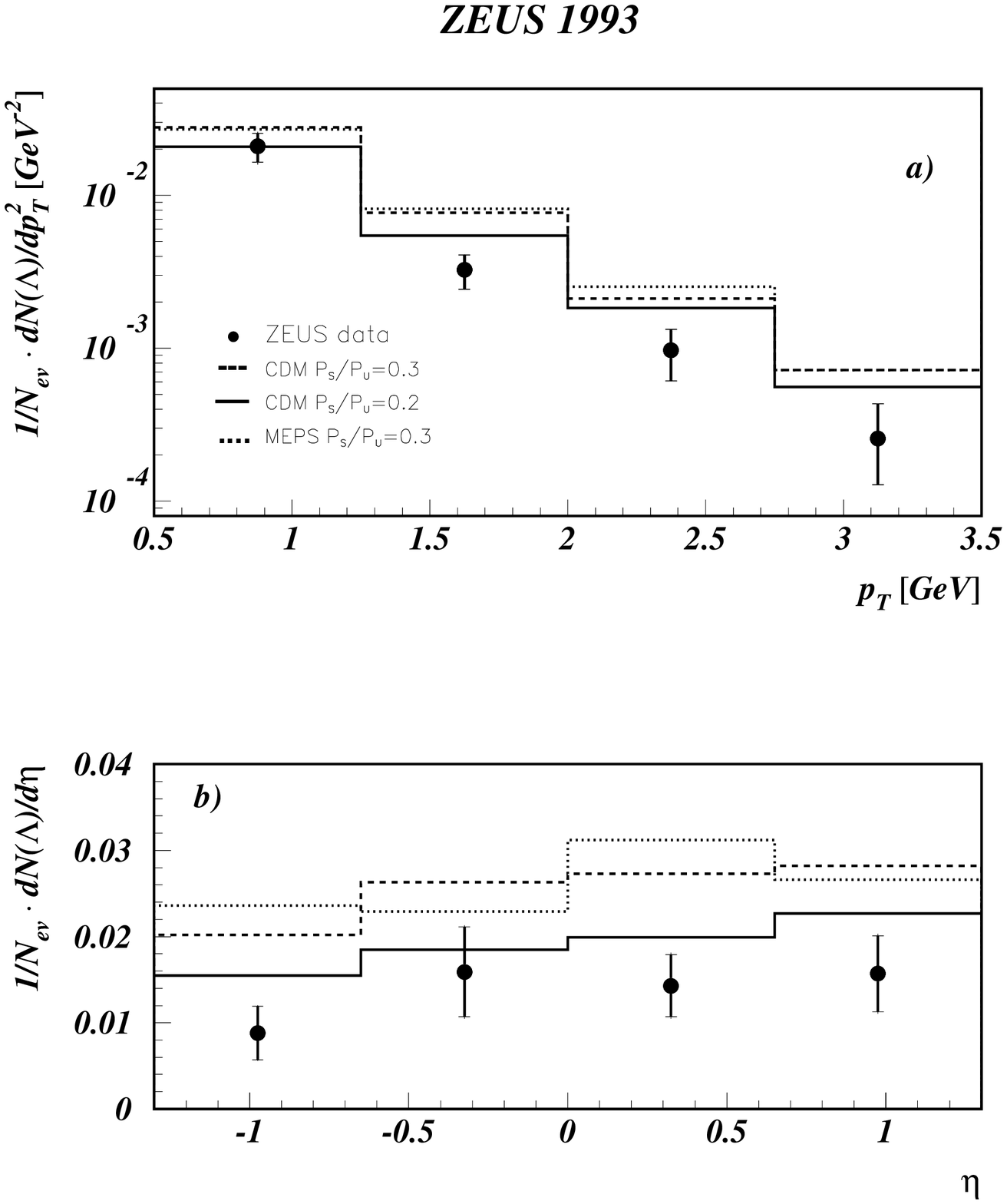}
\caption{  a) Differential multiplicity
of \lam\ versus transverse momentum in the restricted
kinematic \ptr \ and $\eta $ \  range;
b) same for the pseudorapidity of the \lam 's.
The inner error bars show the statistical error and the outer ones
correspond to the statistical and systematic errors added in quadrature.
Statistical errors dominate.
The predictions of the three models discussed in the text are shown.}
\label{fig:lam_rate_pt_eta_lab}
\end{figure}

Figures~\ref{fig:lam_rate_pt_eta_lab}a,~b show the differential \lam\
multiplicity
as a function of the transverse momentum and the pseudorapidity.
The predictions of the CDM and the MEPS Monte Carlo are also
displayed in Fig.~\ref{fig:lam_rate_pt_eta_lab}.
The two CDM curves correspond to samples generated with
different strange-quark suppression factors $P_s/P_u$.
The number of \lam 's with $0.5 < \ptr < 3.5~$GeV and $| \eta | < 1.3 $
produced per event is $0.038 \pm 0.006 \pm 0.002$ for events with
$10<Q^2<640$~GeV$^2$, \ $0.0003<x<0.01$, $y>0.04$.
The measured slope of the \ptr\ distribution of the
\lam\ is $-1.4~\pm~0.3~\pm~0.1$~GeV$^{-1}$,
which, due to the large
statistical uncertainty, is still in agreement with the model predictions.
These values, together with the predictions of the models are listed in
Tab.~\ref{tab:lam_results}.
Monte Carlo studies predict that 16\% (25\%) of the total number of \lam's
(${\overline {\lam}}$'s) will be inside this restricted \ptr\ and $\eta$
region.

The measured \k \ and \lam \ multiplicities seem to be better
described by a model with a strangeness suppression factor of 0.2.

\begin{table}[htbp]
\begin{center}
\begin{tabular}{|l|l|l|}
\hline
     & N(\lam ) / event  & \ptr\ slope [GeV$^{-1}$] \\
\hline
Data    & 0.038 \pms~0.006 \pms~0.002 & --1.4 \pms~0.3 \pms~0.1 \\
CDM &                         &                    \\
\ \ \ with $P_s/P_u = 0.3$  & 0.066 \pms~0.003     & --1.04 \pms~0.07   \\
\ \ \ with $P_s/P_u = 0.2$  & 0.050 \pms~0.002     & --1.00 \pms~0.06    \\
MEPS   &                         &                  \\
\ \ \ with $P_s/P_u = 0.3$  & 0.068 \pms~0.003     & --0.98 \pms~0.06   \\
\hline
\end{tabular}
\caption{
Results of the \lam\  measurement for events with
$10<Q^2<640$~GeV$^2$,
$0.0003<x<0.01$, $y>0.04$ and for a \lam\ with $0.5 < \ptr < 3.5$~GeV,
$ | \eta | < 1.3 $.
The two CDM samples have been generated with a different strange-quark
suppression factor  $P_s/P_u$.
\label{tab:lam_results}
}
\end{center}
\end{table}

\begin{figure}[bp]
\epsfxsize=16cm
\epsfysize=14cm
\epsfbox{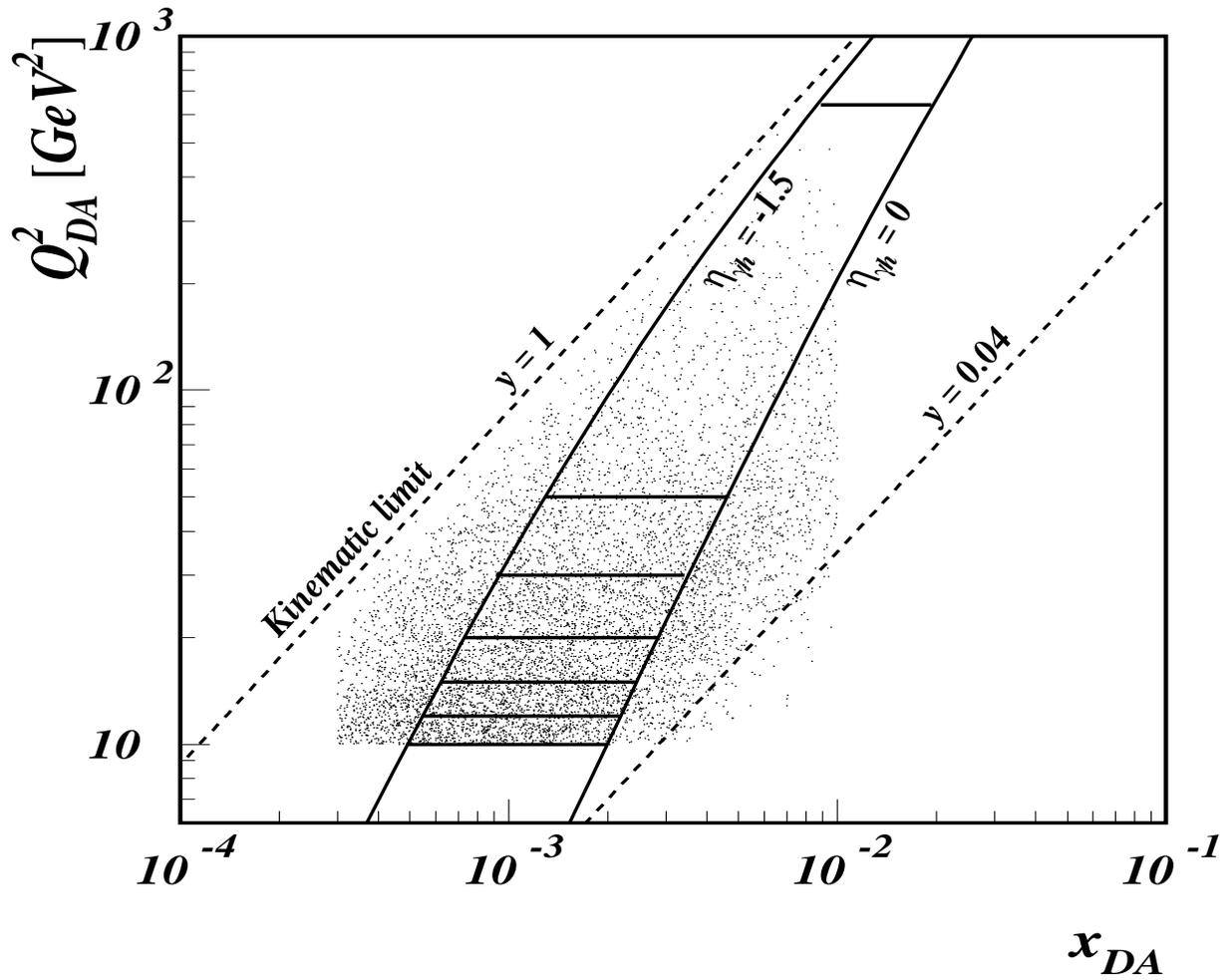}
\caption{The ($x,Q^2$) plane with lines of constant
$\eta_{\gamma_h}$. For the determination of the $Q^2$ dependence of the
 \k\ multiplicity only events with  --1.5 $< \eta_{\gamma_h} < 0$
 are accepted. The horizontal lines correspond to the $Q^2$ bins chosen.}

\label{fig:etagammah}
\end{figure}

\begin{figure}[bp]
\epsfxsize=16cm
\epsfysize=20cm
\epsfbox{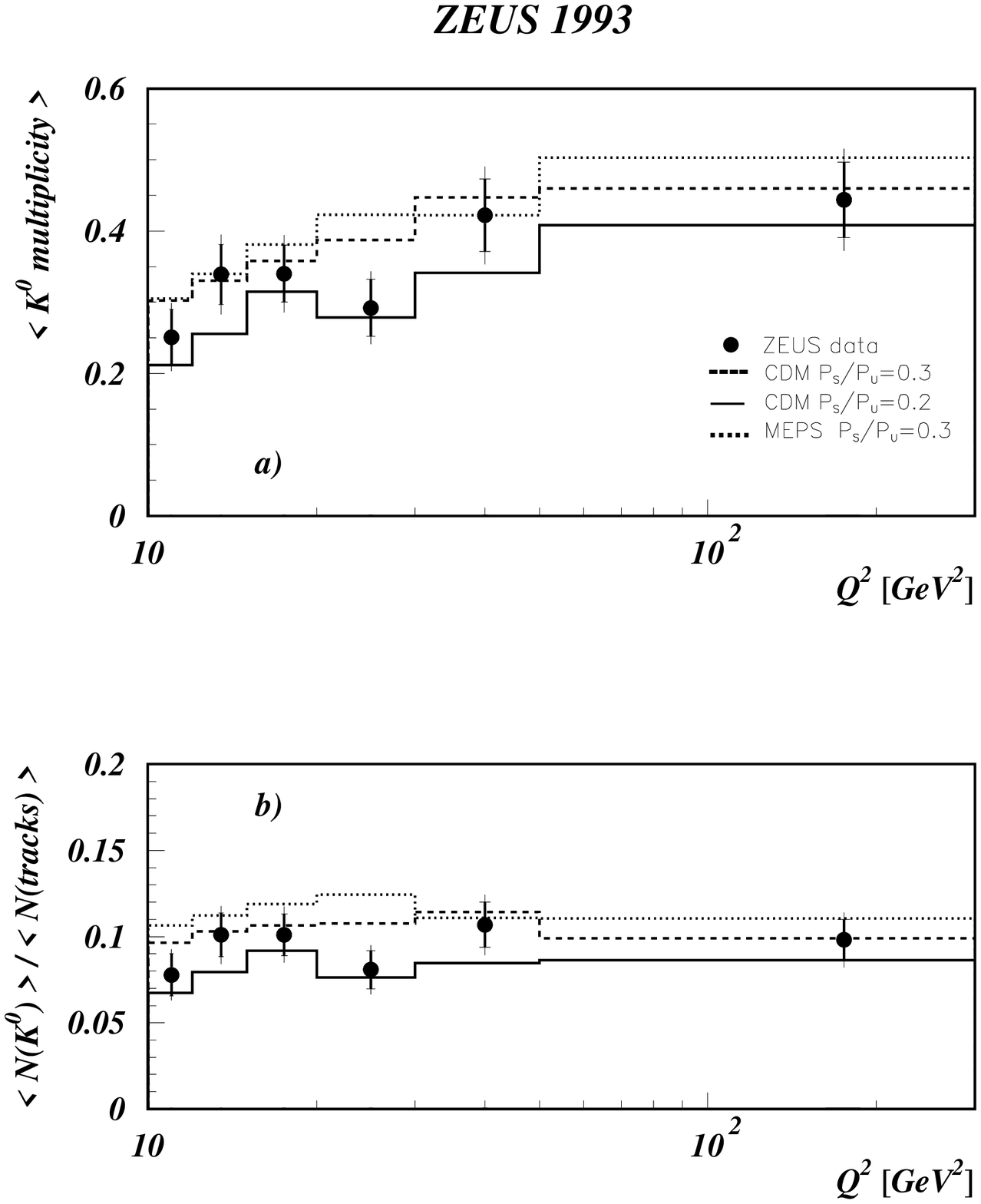}
\caption{
a) Mean multiplicity of \k\ versus the $Q^2$ of the event in the
restricted kinematic ranges. b)
Ratio of the mean \k\ multiplicity to the mean charged particle
multiplicity as a function of the event's $Q^2$.
The inner error bars show the statistical error and the outer ones
correspond to the statistical and systematic errors added in quadrature.
The predictions of three Monte Carlo samples are shown. }
\label{fig:k0_qq}
\end{figure}

We have studied the mean \k\ multiplicity and
the ratio of \k\ to charged particle multiplicities
N(\k )/N(tracks) as a function of the $Q^2$ of the event.
In order to stay in the region of uniform acceptance given
by the inner tracking detector geometry and the analysis cuts, we
restrict this study to events with $-1.5 < \eta_{\gamma_h} < 0$.
Figure~\ref{fig:etagammah} shows the distribution
of our event sample in the ($x,Q^2$) plane. The lines of
constant $\gamma_h$ delimiting the accepted events
and the $Q^2$ bins chosen for this study are shown.
In those bins the variables $Q^2$ and $W$ are correlated:
as $Q^2$ increases from 10 GeV$^2$ to 200 GeV$^2$, the
mean value of $W$ increases from 110 GeV to 160 GeV.
Figure~\ref{fig:k0_qq}a,~b show the mean \k\ multiplicity and
the ratio N(\k )/N(tracks) in the selected bins plotted versus
the mean $Q^2$ of the bins.
The number of charged particles does not include secondary
particles from \k\ and \lam\ decays and from weakly decaying
particles with a lifetime $> 10^{-8}$s.
A slight increase of the $K^0$ multiplicity
and a constant behaviour of N(\k )/N(tracks) are
observed. We have included  the predictions from
the CDM and MEPS Monte Carlo samples, which describe the
data reasonably well.
A study at the Monte Carlo generator level shows that the mean \k\
multiplicity is independent of $Q^2$ for fixed $W$. Since
data and Monte Carlo agree over a wide range of $Q^2$, we conclude
that the mean \k\ multiplicity of our data also shows no $Q^2$
dependence at fixed $W$ within the accuracy of these data.
Furthermore the ratio of \k\ to charged particle multiplicities
is observed to be constant and thus within our experimental errors
this ratio does not depend on the kinematic variables in the region
under study. Therefore we attribute our observed increase of \k\
multiplicity with $Q^2$ to the increase of the corresponding $W$ values.

\subsection{\k\ production in events with a large rapidity gap }

\begin{figure}[htbp]

\includegraphics{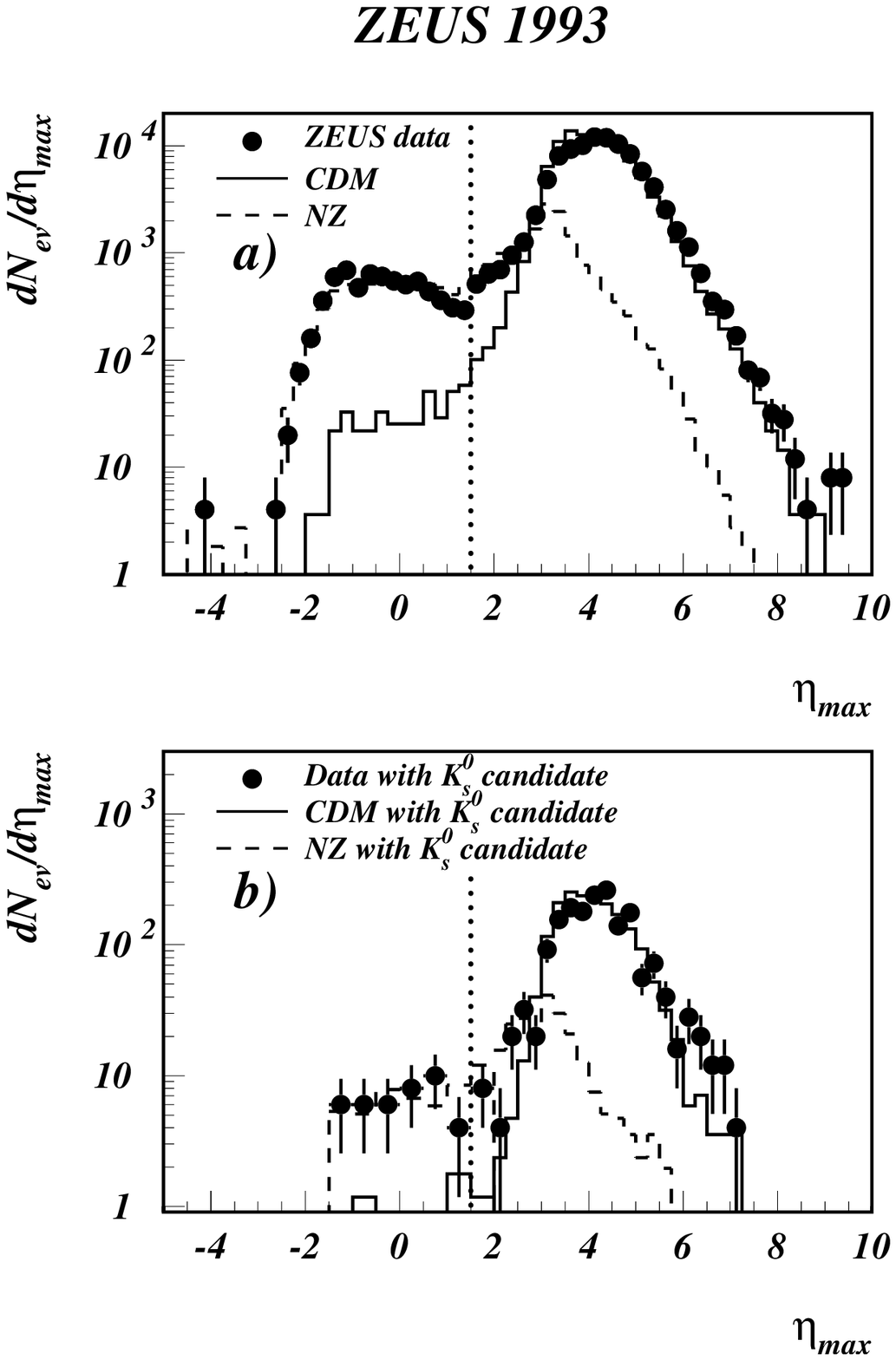}
\unitlength1cm
\begin{picture}(20,20)
\thicklines
\end{picture}

\caption{
a) \etamx\ distribution for all selected events, for one of
   the non-diffractive Monte Carlo samples (CDM) and for one of the
   diffractive Monte Carlo samples (NZ). The relative fractions of
   CDM (88\%) and NZ(12\%) events are chosen so that their sum
   reproduces the data distribution best.
b) \etamx\ distribution of events with a \ks\ candidate in the signal
   region of the $M_{\pi\pi}$ distribution. The predictions of the CDM
   model and of the NZ model are overlaid.
   The dotted line corresponds to \etamx\ $=$ 1.5. }
\label{fig:etamx}
\end{figure}

The DIS data sample is a mixture of non-diffractive and diffractive
 events.
We have searched for differences in \k\ production in these event types.
Following our earlier publications \cite{JETLRG,LUMI},
we separate a non-rapidity gap event sample (NRG) and a LRG event sample
using \etamx . \etamx\ is the largest
pseudorapidity of any calorimeter cluster in an event, where
a cluster is defined as an isolated set of adjacent cells with
summed energy above 400 MeV.
The NRG sample is selected by \etamx\ $>$1.5. It is
dominated by non-diffractive events. The requirement \etamx\ $<$ 1.5
selects a LRG sample which is dominated by diffractive events.
The standard non-diffractive DIS models (CDM, MEPS) give a reasonable
description
of the \etamx\ distribution for values above 1.5 but cannot account
for the excess of events at lower values (see Fig.~\ref{fig:etamx}a).
Values of \etamx\ $>$ 4.3, which are outside the calorimeter
acceptance, occur when energy is deposited in many contiguous cells
around the beam pipe in the proton direction.
An admixture of about 10\% -- 20\% of diffractive events generated with the
NZ or POMPYT Monte Carlo programs to the non-diffractive Monte Carlo sample
gives a reasonable description of the \etamx\ distribution.
The background of non-diffractive DIS events in the LRG sample is
estimated to be 7\% \cite{JETLRG}.
Less than 10\% of the NRG DIS event sample are diffractive events.
Figure~\ref{fig:etamx}b shows the \etamx\ distribution
for those events which have a \ks\ candidate in the signal band.
The \etamx\ distribution of events from one of the non-diffractive (CDM)
and from one of the diffractive (NZ) Monte Carlo samples is also shown.
The excess of \ks\ candidates over predictions from the CDM model
for \etamx\ $<1.5$ represents the
\ks\ production in diffractive events.

As discussed elsewhere \cite{JETLRG,LUMI}, the acceptances for
diffractive events selected by the LRG requirement (\etamx) and for NRG
events are flat with respect to
$W$ and $Q^2$ for $W > 140$~GeV.
We have therefore restricted our comparison to events with
$W > 140$~GeV.   After this additional requirement,
11000 NRG events and 940 LRG events remain.
In the LRG sample we find in the signal region 18 \ks\ candidates over a
background of 2 candidates.

\begin{figure}[tp]
\epsfxsize=16cm
\epsfysize=20cm
\epsfbox{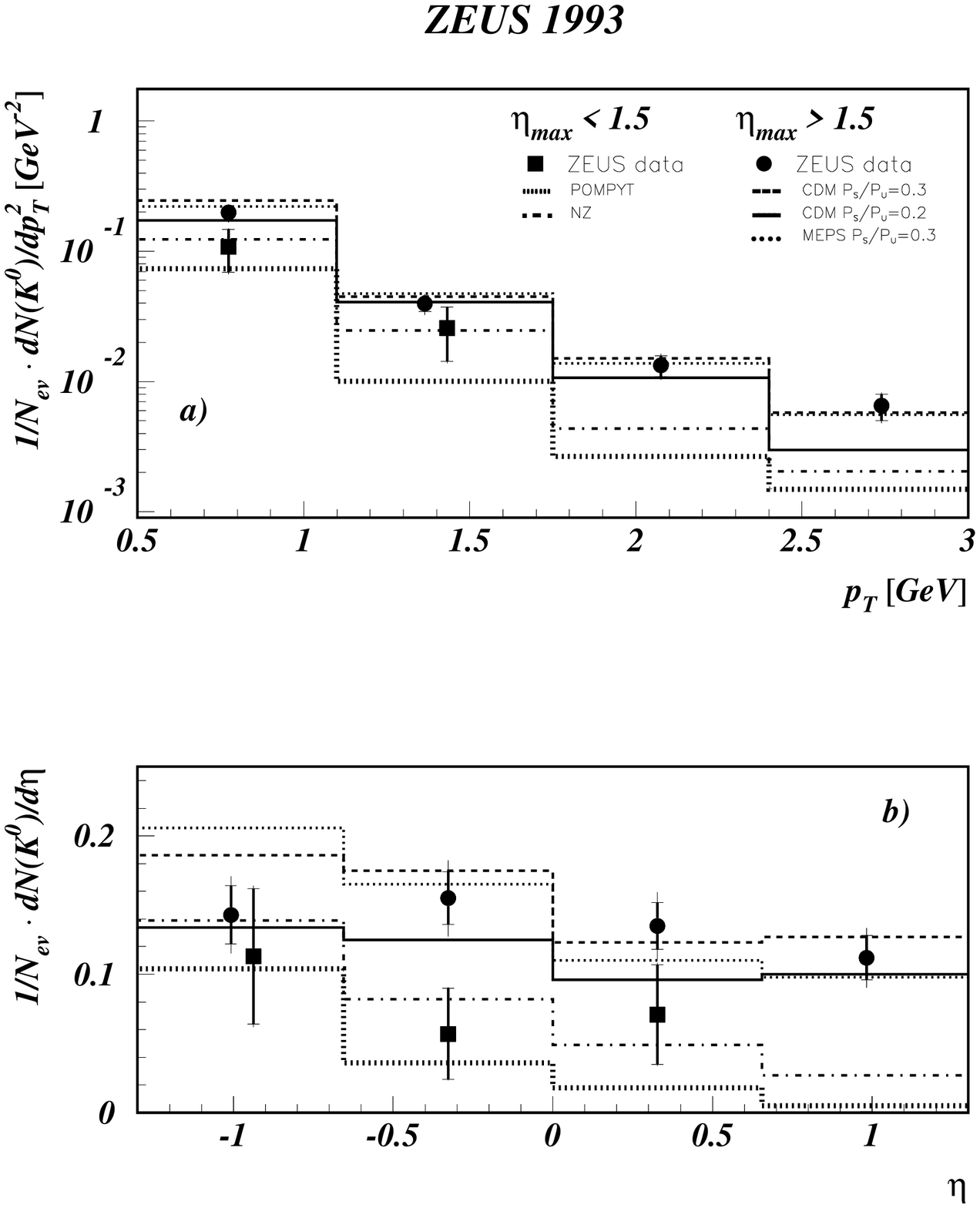}
\caption{ Differential \k\ multiplicity in NRG DIS and LRG DIS events
as a function of their a) transverse momentum; and
b) pseudorapidity.
The inner error bars show the statistical error and the outer ones
correspond to the statistical and systematic errors added in quadrature.
The predictions of five Monte Carlo samples are shown. The two diffractive
Monte Carlo samples are generated with $P_s/P_u = 0.3$.}
\label{fig:k0_rate_pt_eta_lrg}
\end{figure}

Figure~\ref{fig:k0_rate_pt_eta_lrg} shows the
differential \k\ multiplicity as a function of the transverse momentum
and pseudorapidity for NRG and for LRG DIS events separately.
The results in this subsection are not
corrected for either the \etamx\ or the $W$ selection criteria.
The predictions of the standard DIS Monte Carlo programs (CDM and MEPS)
and the diffractive DIS Monte Carlo programs (POMPYT and NZ) are shown.
The \ptr\ distributions  have similar shapes in both data subsamples,
although the multiplicity is lower for the LRG DIS events.
Within the limited statistics of the data, both diffractive models
give a reasonable description of the \k\ multiplicities in LRG events.

Since the invariant mass of the measured hadronic system
in LRG events is smaller than in NRG events, a reduced
\k\ rate is expected in the diffractive events.
We have compared the \k\ multiplicity with the charged
particle multiplicity for both subsamples.
Table~\ref{tab:charged} lists the \k\ multiplicity and the
ratio of the \k\ to charged particle multiplicity
for NRG and LRG DIS events and for the Monte Carlo samples.
If one subtracts the diffractive background, which, as seen from
Fig.~\ref{fig:etamx}, is still present in the NRG DIS sample, the quoted
\k\ multiplicity in the non-diffractive DIS sample increases by 5\%.
The ratios of \k 's to charged tracks for both data samples are
consistent with each other.
Thus, within the limited statistics, these results give no indication
of any additional strange quark enhancement or suppression
in the production mechanism of the LRG final state.

\begin{table}[hptb]
\begin{center}
\begin{tabular}{|c|l|l|l|}
\hline
 & Data type & N(\k ) /event & N(\k ) / N(tracks) \\
\hline
 \etamx\ $>$ 1.5
     & ZEUS data & $0.344\pm 0.023 \pm 0.025$ & $ 0.077 \pm 0.006 \pm 0.008$\\
 NRG & CDM      &                 &                     \\
     & \ \ \ with $P_s/P_u = 0.3$ &  $0.396\pm 0.009$ & $ 0.095 \pm 0.003 $ \\
     & \ \ \ with $P_s/P_u = 0.2$ &  $0.296\pm 0.011$ & $ 0.071 \pm 0.003 $ \\
     & MEPS                       &                   &                     \\
     & \ \ \ with $P_s/P_u = 0.3$ &  $0.375\pm 0.009$ & $ 0.096 \pm 0.003 $ \\
\hline
 \etamx\ $<$1.5
      & ZEUS data & $0.156\pm 0.047 \pm 0.007$ & $ 0.071 \pm 0.021 \pm 0.007$\\
  LRG & POMPYT    & $0.106\pm 0.010$ & $ 0.058 \pm 0.006 $ \\
      & NZ        & $0.173\pm 0.017$ & $ 0.073 \pm 0.007 $ \\
\hline
\end{tabular}
\caption{
The \k\ multiplicity and the ratio of the \k\ and charged particle
multiplicities for NRG and LRG DIS events. The predictions of five
Monte Carlo samples are listed. The diffractive samples are generated
with a strangeness suppression factor $P_s/P_u = 0.3$.
\label{tab:charged}
}
\end{center}
\end{table}

\section{Study of systematic errors}

We have investigated several sources of systematic errors
for our measurements of the \k\ and \lam\ production rates.

1) The sensitivity of the results with respect to the track and primary
 vertex reconstruction methods was determined by repeating the analysis
 with a modified version of the reconstruction package.
 The differences seen are at the 5\% level for the multiplicity
 distributions. No systematic effect is apparent.
 The ratio of \k\ to charged particle multiplicity is similarly
 unaffected.

2) The sensitivity of the results on the choice of the \ks\ and \lam\
 selection criteria has been investigated by varying them by
 $\pm 25$\% of their nominal values.
 The uncertainty in the results from the DIS event selection
 was determined by repeating the analysis with different electron
 finding algorithms and by varying the event
 selection criteria by reasonable values.
 The systematic error from those sources is about
 5\% except for the highest $\eta$ and \ptr\ points in the
 multiplicity distributions and for the results of the LRG event analysis,
 where the error is up to 15\%.
 The mean particle multiplicities per event show lower systematic errors (3\%)
 than the bin by bin errors in the figures.

3) Uncertainties from events rejected by the DIS event selection criteria
   and event migration
 effects were determined by detailed Monte Carlo
 studies of the \k\ and \lam\
 production in the events migrating into and out of the
 selected $Q^2,x,y$ range.
 The \k\ and \lam\ rate of events migrating into this range is
 comparable to that of events migrating out.
 The uncertainty from these sources is at the 5\% level.
 The additional kinematic
 restriction of $W>140$~GeV for the LRG comparison introduces
 a higher uncertainty (7\%) for the  results.
 The mean particle multiplicities show a 2\% uncertainty for NRG
 DIS events and 5\% for the LRG DIS events.

4) We determined a photoproduction contamination in the event sample of
 2.5\%. The event sample which was kinematically restricted to
 $W > 140$~GeV
 contains a higher background of 3.5\% as shown in \cite{F2,Z93F2}.
 We have estimated
 how these photoproduction events affect our analysis
 by studying the stability of the results when varying the
 scattered electron energy and the $\delta$ selection criterion.
 We quote an uncertainty from this source of 3\%.
 The influence on the results from initial state radiative
 events not removed by the $\delta$ requirement is below 3\%
 except for the lowest $\eta$ point in
 Fig.~\ref{fig:k0_rate_pt_eta_lab}b where it is 15\%.

5)
  The \k\ multiplicity versus $Q^2$ is rather sensitive to the
 background below the $M_{\pi\pi}$ signal.
 The combinatorial background increases with $Q^2$ due to the
 observed higher particle multiplicity in events with higher $Q^2$.
 Also migration effects are non-negligible.
 Both effects together may induce
 variations to the  measured values between --11\%
 and +3\% depending on the $Q^2$ bin
 and on the Monte Carlo simulations used to determine them.
 We include an overall systematic error of 10\% to our results
 from these sources.

6)
 The results for the ratio of the \k\ multiplicity to the charged
 particle multiplicity are affected by uncertainties similar to those
 for the \k\ multiplicity alone.
 The variations resulting from different correction
 procedures of calculating the mean charged track multiplicity or from
 using different Monte Carlo samples for the correction are
 within a few percent.
 The relative changes of the ratios of \k\ to  charged
 particle multiplicities for
 the NRG and the LRG data samples are below 5\%
 when different \ptr\ ranges for the charged particles are considered.

7) The strange quark density of the proton structure function
   does not affect our acceptance corrections.

\section{Summary and discussion}

We have measured the \k\ and \lam\ multiplicities
for deep inelastic $ep$ scattering events at $\sqrt{s}=296$~GeV
 with $10~\GeV^2<Q^2<640~\GeV^2$, \ \ $0.0003<x<0.01$ and
$y > 0.04$ \ in the ZEUS experiment at HERA.
We have restricted the analysis to the \k\ and \lam\ kinematic
region \ptr\ $> 0.5 $ GeV and $\left| \eta \right| < 1.3$.
About 23\% (20\%) of the \k\ (\lam )  are predicted to be produced
within this kinematic range.

In this kinematic range the mean number of \k\ (\lam ) per event
is  0.289~\pms~0.015~\pms~0.014 \ (0.038 \pms~0.006 \pms~0.002).
The results on particle production from lower energy $e^+e^-$ data,
which are incorporated in the current DIS Monte Carlo simulation programs
(i.e., strange quark suppression factor $P_s/P_u = 0.3$),
predict  higher \k\ and \lam\ multiplicities than those
observed in the data.
Using a smaller value of 0.2 reduces
the predicted  multiplicity and gives a better agreement with the data,
especially for \lam\ production.
Nevertheless, with $P_s/P_u=0.2$ the prediction for the \lam\
multiplicity is still higher, while the prediction for the \k\ multiplicity
is lower than  the measured values.
The Monte Carlo models allow an adjustment of the production rates of
the different particle types by changing other parameters,
like the ratio of diquarks to single
quarks created from the sea, $P_{qq}/P_q$, as well as the suppression
factor for strange diquarks, $(P_{us}/P_{ud})/(P_s/P_d)$.
Our results indicate the need for tuning these parameters which requires
a detailed measurement of the ratios of pions, kaons, lambdas and protons
over a larger kinematic range. This is beyond the scope of this paper.
The shapes of the distributions for \k 's and \lam 's are
described by both models and do not depend on the chosen parameter $P_s/P_u$.

The mean $K^0$ multiplicity of our data shows no indication for
a $Q^2$ dependence at fixed $W$. Also,
the ratio of \k\ to charged particles is observed to be
independent of the kinematic variables in the range studied.

We observe \k\ production in DIS events with a large rapidity gap
with respect to the proton direction. The \k\
multiplicity in LRG events is approximately a factor
of two lower than in non-diffractive  DIS events.
The ratio of \k\ to charged particles is found to be
the same in both samples.
Thus we observe no additional enhancement or suppression
of neutral kaon production in events with a large rapidity gap
compared to events without a gap.

\section*{Acknowledgements}

The strong support and encouragement by the DESY Directorate have been
invaluable.
The experiment was made possible by the inventiveness
and diligent efforts of the HERA machine group who continued to run HERA
most efficiently during 1993.

The design, construction and installation of
the ZEUS detector have been made by the ingenuity and dedicated efforts of
many people from the home institutes who are not listed here. Their
contributions are acknowledged with great appreciation.
We also gratefully acknowledge the support of the DESY computing and
network services.

\clearpage

\clearpage

\clearpage

\clearpage

\end{document}